\pdfoutput=1

\documentclass[5p,times]{elsarticle}

\usepackage{lineno,hyperref}
\usepackage{amsmath,amssymb,amsfonts}
\usepackage{algorithmic}
\usepackage{graphicx}
\usepackage{textcomp}
\usepackage{xcolor}
\usepackage{caption}
\usepackage{subcaption}
\usepackage{float}
\usepackage{tikz}

\usepackage{filecontents}
\usepackage[labelfont=bf]{caption}
\usepackage{breakcites}
\usepackage{lineno}
\usepackage[linesnumbered,ruled,vlined,onelanguage]{algorithm2e}
\usepackage{hyphenat}
\usepackage{subcaption}
\usepackage{etoolbox}
\usepackage{verbatim}
\usepackage{pdfpages}
\usepackage{lettrine}

\modulolinenumbers[5]

\journal{Journal of \LaTeX\ Templates}

\begin{document}

\begin{frontmatter}

\title{Inspection-L: Self-Supervised GNN Node Embeddings \\ for Money Laundering Detection in Bitcoin}


\author[1]{Wai Weng Lo\corref{cor1}}
\cortext[cor1]{Corresponding author}
\ead{w.w.lo@uq.net.au}

\author[1]{Gayan K. Kulatilleke}
\ead{g.kulatilleke@uq.net.au}

\author[1]{Mohanad Sarhan}
\ead{m.sarhan@uq.net.au}

\author[1]{Siamak Layeghy}
\ead{siamak.layeghy@uq.net.au}

\author[1]{Marius Portmann}
\ead{marius@itee.uq.edu.au}

%l, Application of gradient boosting algo-rithms for anti-money laundering in cryptocurre

\address[1]{The University of Queensland, Brisbane, Australia}

\begin{abstract}
Criminals have become increasingly experienced in using cryptocurrencies, such as Bitcoin, for money laundering. The use of cryptocurrencies can hide criminal identities and transfer hundreds of millions of dollars of dirty funds through their criminal digital wallets. However, this is considered a paradox because cryptocurrencies are goldmines for open-source intelligence, giving law enforcement agencies more power when conducting forensic analyses. This paper proposed Inspection-L, a graph neural network (GNN) framework based on a self-supervised Deep Graph Infomax (DGI) and Graph Isomorphism Network (GIN), with supervised learning algorithms, namely Random Forest (RF), to detect illicit transactions for anti-money laundering (AML). To the best of our knowledge, our proposal is the first to apply self-supervised GNNs to the problem of AML in Bitcoin. The proposed method was evaluated on the Elliptic dataset and shows that our approach outperforms the state-of-the-art in terms of key classification metrics, which demonstrates the potential of self-supervised GNN in the detection of illicit cryptocurrency transactions.
\end{abstract}


\end{frontmatter}

\section{Introduction}
The advent of the first cryptocurrency---Bitcoin \cite{Nakamoto}---has revolutionized the conventional financial ecosystem, as it enables low-cost, near-anonymous, peer-to-peer cash transfers within and across various borders. Due to its pseudonymity,  many cybercriminals, terrorists, and hackers have started to use cryptocurrency for illegal transactions. For example, the WannaCry ransomware attack used Bitcoin \cite{kshetri2017crypto} as the payment method due to its non-traceability. The criminals received nearly 3.4 million (46.4 BTC) within four days of the WannaCry attack \cite{kshetri2017crypto}. 
Therefore, effective detection of illicit transactions in Bitcoin transaction graphs is essential for preventing illegal transactions.
%please check intended meaning
Paradoxically, cryptocurrencies are goldmines for open-source intelligence, as transaction network data are publicly available, enabling law enforcement agencies to conduct a forensic analysis of the transaction's linkages and flows. 
However, the problem is challenging for law enforcement agencies, owing to its volume \footnote{As of 2022 Aug 09, the volume of the entire BTC transaction record, the blockchain, is 420GB, with an average growth rate of 129\%}, the untraceable p2p cross-border nature of Bitcoin transactions, and the use of technologies such as mixers and tumblers. 

Graph representation learning has shown great potential for detecting money laundering activities using cryptocurrencies. GNNs are tailored to applications with graph-structured data, such as the social sciences, chemistry, and telecommunications, and can leverage the inherent structure of the graph data by building relational inductive biases into the deep learning architecture. This provides the ability to learn, reason, and generalize from the graph data, inspired by the concept of message propagation~\cite{gnn_suv}. 

The Bitcoin transaction flow data can naturally be represented in graph format. A graph is constructed from the raw Bitcoin data and labeled such that the nodes represent transactions and the edges represent the flow of Bitcoin currency (BTC) from one transaction to the next in the adjacency matrix. Both the topological information and the information contained in the node features are crucial for detecting illicit transactions.

This paper proposes Inspection-L, a Graph Neural Network (GNN) framework based on an enhanced self-supervised Deep Graph Infomax (DGI) \cite{dgi} and supervised Random Forest (RF)-based classifier to detect illicit transactions for AML. 

Specifically, we investigate the Elliptic dataset \cite{AML}, a realistic, partially labeled Bitcoin temporal graph-based transaction dataset consisting of real entities belonging to licit (e.g., wallet, miners), illicit entities (e.g., scams, terrorist organizations, ransomware), and unknown transaction categories. The proposed Inspection-L framework aims to detect illegal transactions based on graph representation learning in a self-supervised manner. Current graph machine learning approaches, such as \cite{AML}, generally apply supervised graph neural network approaches to the detection of illicit transactions. However, supervised learning requires manual labeling. In the AML scenario, building an effective model that utilizes unknown label data is required, since human's labeling Bitcoin data could be costly and ineffective. %please check intended meaning
It also only performs well when the labels are enough. Thus, exploiting unlabeled data to improve performance is critical for AML. On the other hand, self-supervised graph neural network algorithms \cite{liu2021self}\cite{liu2021graph} allow for the unknown label data to be exploited, which can improve the quality of representation for the downstream tasks such as fraud transaction detection in Bitcoin. Furthermore, in supervised learning, GNN is limited to capturing K-hop neighbor information; for example, once the hops of the neighbor are larger than k, the supervised learning GNN fails to capture that node information.

In this paper, we applied DGI self-supervised learning to capture the global graph information, as this is not limited to capturing the K-layer neighborhood information, where every node can access the entire graph's structural pattern and node information using random shuffle node features. The DGI discriminator tries to determine wherever the node feature is shuffled or not.  %please check intended meaning
Thus, every node can access global parts of the node's properties, rather than K-layer neighborhood information. 

We demonstrate how the self-supervised DGI algorithm can be integrated with standard machine learning classification algorithms, i.e., Random Forest, to build an efficient anti-money-laundering detection system. We show that our Inspection-L method outperforms the state-of-the-art in terms of F1 score.

In summary, the key contributions of this paper are:

\begin{itemize}
\setlength\itemsep{0.8em}

\item  Different from most existing works, which typically use supervised graph representation learning to generate node embeddings for illegal transaction detection, we use a self-supervised learning approach to learn the node embeddings without using any labels.

\item The proposed Inspection-L is based on a self-supervised DGI combined with the Random Forest (RF) supervised machine learning algorithms, to capture topological information and node features in the transaction graph to detect illegal transactions. To the best of our knowledge, our proposal is the first to utilize self-supervised GNNs to generate node embeddings for AML in Bitcoin.
\item The comprehensive evaluation of the proposed framework using the Elliptic benchmark datasets demonstrates superior performance compared to other, supervised machine learning approaches.
\end{itemize}

\section{\uppercase{Related Works}}
% \subsection{Android Malware Classification based on Machine Learning}
Mark et al. \cite{AML} created and published the Elliptic dataset, a temporal graph-based Bitcoin transaction dataset consisting of over 200K Bitcoin node transactions, 234K payment edges, and 49 transaction graphs with distinct time steps. Each of the transaction nodes was labeled as a "licit", "illicit", or "unknown" entity. They evaluated the Elliptic dataset using various machine learning methods, including Logistic Regression (LR), Random Forest (RF), Multilayer Perceptrons (MLP) \cite{bishop2006pattern}, Graph Convolutional Networks (GCNs) \cite{kipf2016semi}  and EvolveGCN \cite{pareja2020evolvegcn}. They retrieved a recall score in the illicit category of 0.67 using RF and 0.51 using GCNs.

Yining et al. \cite{hu2019characterizing} collected the Bitcoin transaction graph data between July 2014 and May 2017 by running a Bitcoin client and used an external trusted source, "Wallet Explorer", a website that tracks Bitcoin wallets, to label the data. They first highlighted the differences between money laundering and regular transactions using network centrality such as PageRank, clustering coefficient \cite{bondy1976graph}, then used a node2vec-based \cite{grover2016node2vec} classifier to classify money laundering transactions. The research also indicated that statistical information, such as in-degree/out-degree, number of weakly connected components, and sum/mean/standard deviation of the output values, could distinguish money laundering transactions from legal transactions. However, this approach only considers graph topological patterns, without considering node features.  Vassallo et al. \cite{vassallo2021application} focused on the detection of illicit cryptocurrency activities (e.g., scams, terrorism financing, and Ponzi schemes). Their proposed detection framework is based on Adaptive Stacked eXtreme Gradient Boosting (ASXGB), an enhanced variation of eXtreme
Gradient Boosting (XGBoost). ASXGB was evaluated using the Elliptic dataset, and the results demonstrate its superiority at both the account and transaction levels.

Chaehyeon et al. \cite{lee2019toward} applied supervised machine learning algorithms to classify illicit nodes in the Bitcoin network. They used two supervised machine learning models, namely, Random Forest (RF) and Artificial Neural Network (ANN) \cite{bishop2006pattern} to detect illegal transactions. First, they collected the legal and illegal Bitcoin data from the forum sites "Wallet Explorer" and "Blockchain Explorer". Next, they performed feature extraction based on the characteristics of Bitcoin transactions, such as transaction fees and transaction size. The extracted features were labeled legal or illegal for supervised training. 
The results indicated that relatively high F1 scores could be achieved; specifically, ANN and RF achieved $0.89$ and $0.98$ F1 scores, respectively. In \cite{Alarab} proposed using GCNs intertwined with linear layers to classify illicit nodes of the Elliptic dataset \cite{AML}. An overall classification accuracy and recall of 97.40\% and 0.67, respectively, can be achieved to detect illicit transactions. In \cite{nan2018bitcoin}, the authors used an autoencoder with graph embedding to detect mixing and demixing services for Bitcoin cryptocurrency. They first applied graph node embedding to generate the node representation; then, a K-means algorithm was applied to cluster the node embeddings to detect mixing and demixing services. The proposed model was evaluated based on real-world Bitcoin datasets to evaluate the model's effectiveness, and the results demonstrate that the proposed model can effectively perform demix/mixing service anomaly detection.

Lorenz et al. \cite{lorenz2020machine} proposed active learning techniques by using a minimum number of labels to achieve a high rate of detection of illicit transactions on the Elliptic dataset. In \cite{pham2016anomaly}, the authors applied unsupervised learning to detect suspicious nodes in the Bitcoin transaction graph. They used various kinds of unsupervised machine learning algorithms, such as K-means and Gaussian Mixture models, to cluster normal and illicit nodes. 
However, since the Bitcoin transaction dataset they used lacked ground-truth labels, they simply used the internal index to validate the clustering algorithm, without confirming that those nodes are actually malicious transactions. 
Monamo et al. \cite{monamo2016unsupervised} applied trimmed-Kmeans to detect fraud in the Bitcoin network. They used various graph centrality measures (i.e. in degree, out-degree of the Bitcoin transactions) and currency features (i.e. the total amount sent), which were then used for Bitcoin transaction clustering. 
However, similar to \cite{pham2016anomaly}, due to the unavailability of ground-truth labels, they used clustering performance metrics such as "within the sum of squares", without being able to validate the true nature of the Bitcoin transaction anomalies. Shucheng et al. \cite{li2021self} proposed SIEGE, a self-supervised graph learning approach for Ethereum phishing scam detection, using two pretext tasks to generate node embeddings without using labels and an incremental paradigm to capture data distribution changes for over half a year. However, a significant limitation of this approach is that is does not consider the Bitcoin context and is limited to detecting Ethereum phishing scams. Additionally, their simple application of GCNs\cite{kipf2016semi} in the pretext task phase is much less effective than the Weisfeiler--Lehman (1-WL) test\cite{shervashidze2011weisfeiler}.  

In contrast with related studies, our approach can detect not only phishing scams but also other illicit transactions, such as terrorist organizations, ransomware and Ponzi schemes, by utilizing the Elliptic dataset \cite{AML}.

\section{\uppercase{Background}}

\begin{figure}[h]

    \centering
        \includegraphics[width=0.8\columnwidth]{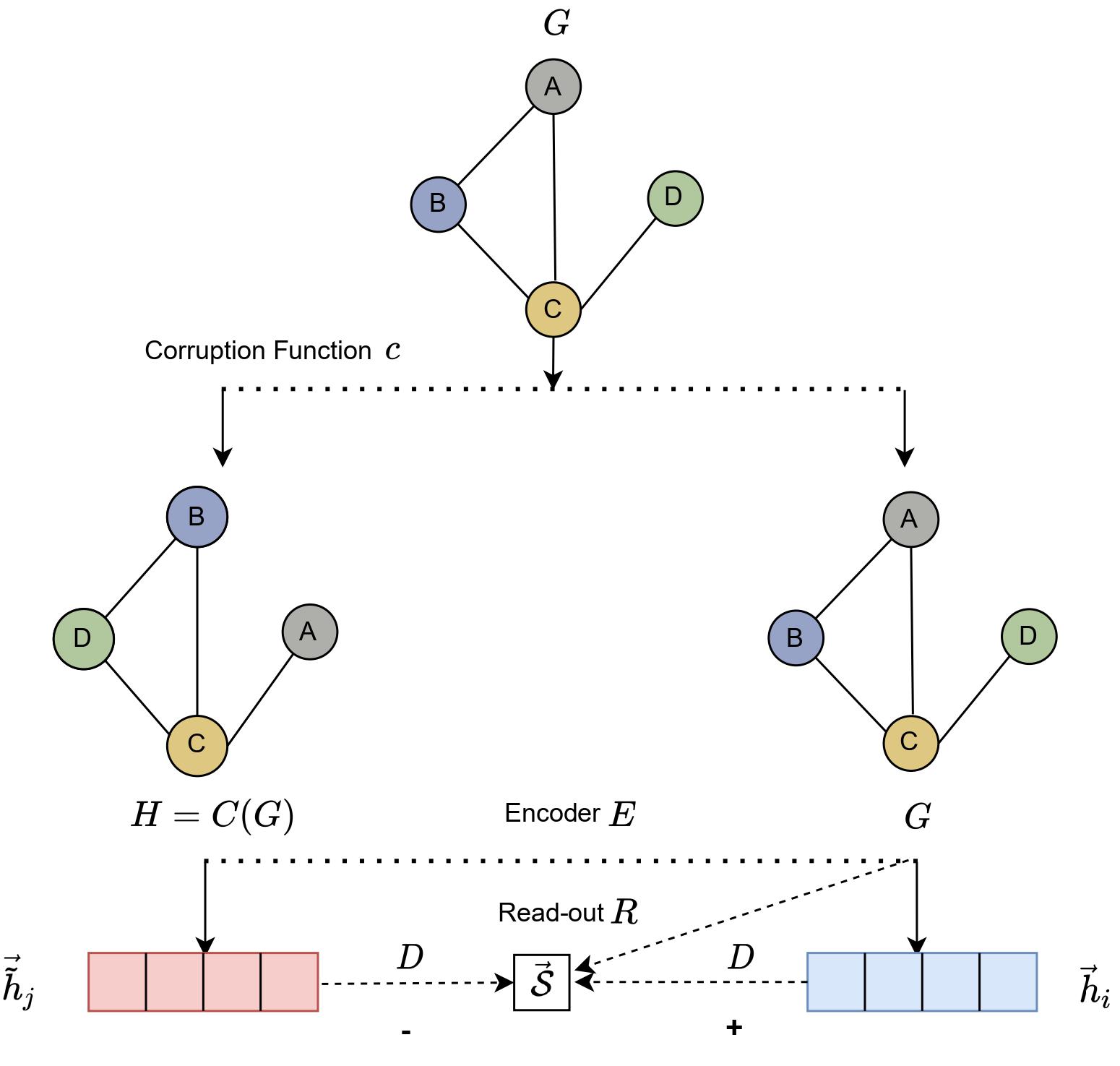}
        %\label{rfidtest_yaxis}
    \caption{Overview of Deep Graph Infomax }
    \label{fig:dgi_overview}
\end{figure}

The main innovation of our proposed model is its use of DGI \cite{dgi} with our proposed GIN encoder to learn node embeddings in a self-supervised manner. Then, the node embeddings can be treated as enhanced features and be combined with the raw features for standard supervised RF machine learning algorithms to classify illicit transaction. %please check intended meaning
 This has a clear advantage over simple features, as inputs to overall graph-structured patterns are available for the downstream classifier.   %please check intended meaning

Consequently, the current graph-based approaches \cite{AML} \cite{Alarab} try to apply a supervised GCN-based approach to capture the overall graph-structured patterns. However, the main limitation is that GCN can only capture the neighborhood information of limited K layers, not the global view graph and node information, due to the threat of overfitting.
While some models, such as FDGATII \cite{kulatilleke2021fdgatii}, are capable of a larger K, these are still limited by their layer structure and finite k.
On the other hand, our Inspection-L approach allows for every node to obtain access to the structural patterns of the entire graph, which can capture more global neighborhood information. The proposed method considers that the message-passing functions of \cite{AML} \cite{Alarab} are not powerful enough, as they lack injective functions. Therefore, we proposed a GIN encoder to make the message propagation function more robust.

\subsection{\textbf{Graph Neural Networks}}
GNNs is a deep learning approach for graph-based data and a recent and highly promising area of machine learning \cite{kulatilleke2021fdgatii}. The key feature of GNNs is their ability to combine a topological graph structure with features. For each node in a graph, this means aggregating neighboring node features to leverage a new representation of the current node that considers the neighboring information. The output of this process is known as embeddings. Final node embeddings are low- or n-dimensional vector representations that capture topological and node properties. 
%This process is able to be performed for edges and graphs in a similar fashion to output edge and graph embeddings. 
Embeddings can be learned in a supervised or unsupervised manner and used for downstream tasks such as node classification, clustering, and link prediction \cite{kulatilleke2021fdgatii}. 
The $k$-th layer of a typical GCN is:  
\begin{equation}
%Z^{l+1}=\sigma\left(X^{l} W_{0}^{l}+\tilde{A} X^{l} W_{1}^{l}\right)
h_v^{(k)} =   \sigma \left(W \cdot {\rm MEAN} \left\{h_u^{(k-1)},\ \forall u \in {N}(v) \cup \{v\} \right\} \right).
\end{equation}

where $h_v^{(k)}$ is the feature vector of node $v$ at the $k$-th iteration/layer, $h_v^{(0)} = X_v$, and ${N}(v)$ is the set of neighbor nodes of $v$. 
$W^{(l)}$ is the weight matrix that will be learned for the downstream tasks. $\sigma$ is an activation function, typically ReLU, for computing node representations. 
%Let there be $L$ layers of graph convolutions, the output $Z(L)$ is the matrix consisting of all node embeddings after $L$ layer transformation.

\subsection{\textbf{Graph Isomorphism Network}}

Graph Isomorphism Network (GIN) is theoretically a maximally powerful GNN proposed by Xu et al. \cite{xu2018powerful}. The main difference between GIN and other GNNs is the message aggregation function, which is shown below:
\begin{equation}
h_{v}^{(k)}=\mathrm{MLP}^{(k)}\left(\left(1+\epsilon^{(k)}\right) \cdot h_{v}^{(k-1)}+\sum_{u \in {N}(v)} h_{u}^{(k-1)}\right)
\label{eq:gin}
\end{equation}

GCNs is less effective than the Weisfeiler--Lehman (1-WL) \cite{shervashidze2011weisfeiler} test due to the single-layer aggregation function, which is same as the hash function of a 1-WL algorithm. According to \cite{xu2018powerful}, a single, non-linear layer is insufficient for graph learning. Thus, GCN message passing functions are not necessarily injective. Therefore, GIN \cite{xu2018powerful} was proposed to make the passing function injective, as shown in Equation \ref{eq:gin}, where $\varepsilon^{(k)}$ is a scalar parameter, and MLP stands for multilayer perceptron. ${h}_{v}^{(k)} \in \mathbb{R}^{d}$ is the embedding of node $v_{i}$ at the $k$-th layer, ${h}_{v}^{(0)}={x}_{v}$ is the original input node features, and $N\left(v_{i}\right)$ is the set of neighboring nodes of node $v_{i}$.  We can stack $k$ layers to obtain the final node representation ${h}_{v}^{(k)}$.

\subsection{\textbf{Deep Graph Infomax}}
Deep Graph Infomax (DGI) \cite{dgi} is a self-supervised graph representation learning approach that relies on maximizing the mutual information between patch representations and the global graph summary. The patch representations summarize subgraphs, allowing for the preservation of similarities at the patch level. A trained encoder in DGI can be reused to generate node embeddings for downstream tasks, such as node clustering. 

Most of the previous works on self-supervised representation learning approaches rely on the random walk strategy \cite{Hamilton2017}\cite{zhang2019heterogeneous}, which is extremely computationally expensive because the number of walks depends on the number of nodes on the graph, making it unscalable for large graphs. Moreover, the choice of hyperparameters (length of the walk, number of walks) can significantly impact the model performance. Overall, DGI does not require supervision or random walk techniques. Instead, it guides the model to learn node connections by simultaneously leveraging local and global information in a graph \cite{dgi}. 
%Therefore, it can be used to train node embeddings in a self-supervised manner.

Figure \ref{fig:dgi_overview} shows the overall operation of DGI. 
%The DGI algorithm operates on the following bases:
%\begin{itemize}
${G}$ is a true graph with the true nodes, the true edges  that connect them, and real node features associated with each node. 
${H}$ is a corrupted graph where the nodes and edges have been changed using a corruption function. \cite{dgi} suggests that the corruption function can randomly shuffle each node feature and maintain the same edges as the true graph ${G}$. 
%\end{itemize}

The DGI training procedure consists of four components:
\begin{itemize}
  \item A corruption procedure ${C}$ that changes the real input graph $G$ into a corrupted graph ${H}= (C(G))$. This can be achieved by randomly shifting the node features among the nodes in a real graph ${G}$ or by adding and removing an edge from the real graph ${G}$.

  \item An encoder ${E}$ that computes the node embeddings of a corrupted graph and a real graph. This can be achieved using various graph representation methods, such as Graph Convolutional Networks (GCNs) \cite{kipf2016semi}, Graph Attention Networks (GATs) \cite{velivckovic2017graph} or Graph Transformer Networks (GTNs) \cite{yun2019graph}.
  \item The node embedding vectors for each node in the real graph are summarized into a single embed vector of the entire graph $\overline{s}$ (global graph summary) by using a read-out function ${R}$ to compute the whole graph embeddings.
  \item A discriminator ${D}$, which is a logistic non-linear sigmoid function, compares a real node embedding vector $\vec{h}_{i}$ and a corrupted node embedding $\widetilde{h}_{i}$  against the whole real graph embedding $\overline{s}$, and provides a score between 0 and 1, as shown in Equation~\ref{eq:dgi}. This binary cross-entropy loss objective function \cite{dgi} can be applied to discriminate between the embedding of the real node and the corrupted node to train the encoder ${E}$.
  
\end{itemize}

\begin{equation}
\begin{split}
{L}=\frac{1}{N+M}\Biggl(\sum_{i=1}^{N} \mathbb{E}_{(\mathbf{X}, \mathbf{A})}\left[\log {D}\left(\vec{h}_{i}, \vec{s}\right)\right]+\\ \\
\sum_{j=1}^{M} \mathbb{E}_{(\overline{\mathbf{X}}, \overline{\mathbf{A}})}\left[\log \left(1-{D}\left(\vec{h}_{j}, \vec{s}\right)\right)\right]\Biggr)
\end{split}
\label{eq:dgi}
\end{equation}

\section{\uppercase{Proposed Method}}

\begin{figure*}[h]
    \raggedleft 
        \includegraphics[width=2.0\columnwidth]{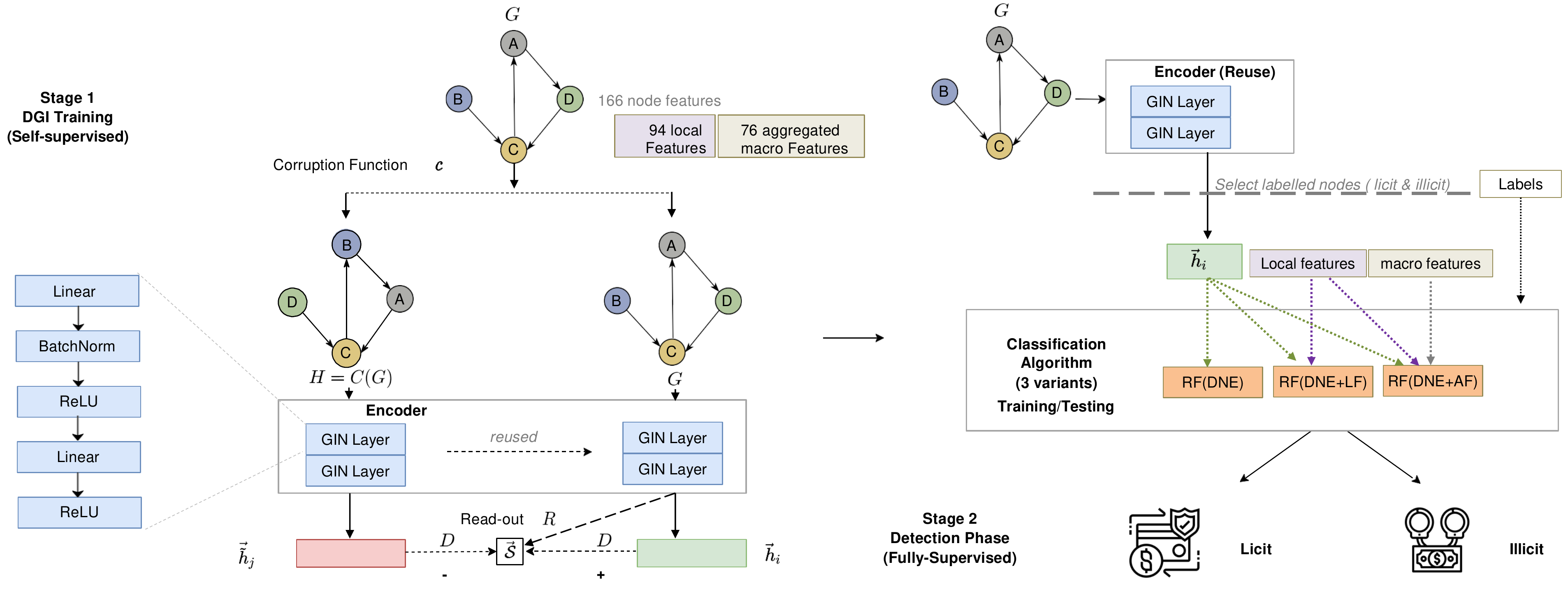}
        %\label{rfidtest_yaxis}
    \caption{Proposed Method}
    \label{fig:proposed_method}
\end{figure*}

%AAAAAAAAAAAAAAAAAAAAAAAAAAAAAAAAAAAAAAAAAAAA
\IncMargin{1.5em}
\begin{algorithm*}[!t]\footnotesize
\SetKwData{This}{this}\SetKwData{Up}{up}
  \SetKwFunction{Union}{Union}\SetKwFunction{FindCompress}{FindCompress}
  \SetKwInOut{Input}{input}\SetKwInOut{Output}{output}

\Indp\Indpp
  \Input{
  Set of training graphs $G^+ =  \{{G}({V}, {A},{X})$\};\\ 
  Number of training epochs  $K$;\\ 
  Corruption function  $C$;\\ 
  All 166 Features ($AF$);\\ 
  First 94 Local Features ($LF$);\\ 

  \vspace{0.05cm}
  }
  
  \Output{
  Optimized GIN encoder $g$,   Optimized RF $h_{-} R $

    }
  \BlankLine
    Initialize the parameters $\theta$ and $\omega$ for the encoder $g$ and the discriminator $D$;
    
\ForEach {batch $G \in G^+ $}{
  \For{$epoch\leftarrow 1\ \KwTo\ K$}{
    $h_{i}=g(G, \theta)$ \\
    $\widetilde{h}_{i}=g(C(G), \theta) $\\
    $\bar{s}= \sigma\left(\frac{1}{n} \sum_{i=1}^{n} h_{i}^{(L)}\right)$
     
    ${D}\left({h}_{i}, \bar{s}\right)=\sigma\left({h}_{i}^{T} \mathbf{w} \bar{s}\right)$\\

    ${D}\left(\widetilde{h}_{i}, \bar{s}\right)=\sigma\left(\widetilde{h}_{i}^{T} \mathbf{w} \bar{s}\right)$ \\ 
    
%\textcolor{blue}{${L}_{D G I}=\frac{1}{N+M}\Biggl(\sum_{i=1}^{N} \mathbb{E}_{(\mathbf{X}, \mathbf{A})}\left[\log {D}\left(\vec{h}_{i}, \vec{s}\right)\right]+
%\sum_{j=1}^{M} \mathbb{E}_{(\overline{\mathbf{X}}, \overline{\mathbf{A}})}\left[\log \left(1-{D}\left(\vec{h}_{j}, \vec{s}\right)\right)\right]\Biggr)$}

\textcolor{black}{${L}_{D G I}=\frac{1}{N + M}\Biggl(\sum_{i=1}^{N} \mathbb{E}_{(\mathbf{X}, \mathbf{A})}\left[\log {D}\left(\vec{h}_{i}, \vec{s}\right)\right]+
\sum_{j=1}^{M} \mathbb{E}_{(\overline{\mathbf{X}}, \overline{\mathbf{A}})}\left[\log \left(1-{D}\left(\vec{h}_{j}, \vec{s}\right)\right)\right]\Biggr)$} \\
    $\theta, \omega \leftarrow \operatorname{Adam}\left({L}_{DGI}\right)$
     
}}

  \BlankLine
    Select  labeled node embedding $h_i$ from $h_{i}=g(G, \theta)$ and corresponding labels $y$ for $G \in \text{training set}$;\\
%$h_{i}=g(G, \theta)$ \\
$h_{-} R \leftarrow \mathrm{RF}((h_{i} || \text{\{$AF$ or $LF$\}}),y )$\\
\Return  $h_{-} R, g$

\caption{Pseudocode for Our Proposed Algorithm}
\label{alg:Inspection-L}
\Indp\Indpp
\end{algorithm*}\DecMargin{1em}

%AAAAAAAAAAAAAAAAAAAAAAAAAAAAAAAAAAAAAAAAAAAA
To construct Bitcoin transaction graphs from the dataset, we used 49 different Bitcoin transaction graphs (TGs) \cite{AML} using time steps so that the nodes can be represented as node transactions and the edges can be represented as flows of Bitcoin transactions. This is a very natural way to represent Bitcoin transactions. 

The pseudocode and overall procedure of our proposed algorithm are shown in Algorithm \ref{alg:Inspection-L} and Figure. \ref{fig:proposed_method}. The proposed framework consists of two-stage: DGI training for node embedding extraction to perform feature augmentation, supervised machine learning classification.

\subsection{DGI Training}\label{DGI}

To train the proposed model, the input includes the transaction graphs ${G}$ with node features (i.e., all 166 features, which is a combination of local and macro features, which we denote AF, or only the 94 local features), and the specified number of training epochs $K$ to extract true node embeddings and corrupted node embeddings. Before this, we need to define the corruption function ${C}$  to generate the corrupted transaction graphs ${C(G)}$ for our GIN encoder to extract the corrupted node embeddings. In this paper, we randomly shuffled all the node features among the nodes in real transaction graphs ${G}$ to generate the corrupted transaction graphs for each real graphs by shuffling the feature matrix in rows $\mathbf{X}$ by using Bernoulli distribution. Overall, instead of adding or removing edges from the adjacency matrix such that $\mathbf{A}_G \neq \mathbf{A}_H$,  we use corruption function ${C}$, which shuffle the node features such that $\mathbf{X}_G \neq \mathbf{X}_H$, and retain the adjacency matrix, i.e., $(\mathbf{A}_G =\mathbf{A}_H)$. Note that the corruption function only changes the node features, and not the structure; therefore, $N_G=N_H$. In case of the DGI implementation, we now have $N=M$.

For each batch of graph data $G$ in the training epoch, in Algorithm~\ref{alg:Inspection-L} from Line  3 to 4, we use our proposed GIN encoder to extract true node and corrupted node embeddings. Our proposed GIN encoder is shown in Figure \ref{fig:proposed_method} with two layers  of MLP, which consists of 128 hidden units, ReLU activation function and Batch normalization (as shown in Algorithm \ref{alg:BN}) \cite{ioffe2015batch}.

The design of the MLPs is motivated by the fundamental goal of a GNN-based model. Ideally, various types of different graph patterns should be distinguishable via the graph encoder, which means that different graph structures should be mapped to different locations in the embedding space. This requires the ability to solve the graph isomorphism problem, where non-isomorphic graphs should be mapped to different representations.

We applied a full neighbor sampling technique and used two-hop neighbor samples for the GIN encoder with Batch normalization, as DGI benefits from employing wider rather than deeper models \cite{dgi}.

%AAAAAAAAAAAAAAAAAAAAAAAAAAAAAAAAAAAAAAAAAAAA

%AAAAAAAAAAAAAAAAAAAAAAAAAAAAAAAAAAAAAAAAAAAA
\IncMargin{1.5em}
\begin{algorithm}[!t]\footnotesize
\SetKwData{This}{this}\SetKwData{Up}{up}
  \SetKwFunction{Union}{Union}\SetKwFunction{FindCompress}{FindCompress}
  \SetKwInOut{Input}{input}\SetKwInOut{Output}{output}

\Indp\Indpp
  \Input{
Values of $x$ over a mini-batch: ${B}=\left\{x_{1 \ldots m}\right\}$ Parameters can be learned: $\gamma, \beta$
  \vspace{0.05cm}
  }
  
  \Output {
$\left\{y_{i}=\mathbf{B N}_{\gamma, \beta}\left(x_{i}\right)\right\}$
}
  \BlankLine

$\mu_{{B}} \leftarrow \frac{1}{m} \sum_{i=1}^{m} x_{i}$  \\

$\sigma_{{B}}^{2} \leftarrow \frac{1}{m} \sum_{i=1}^{m}\left(x_{i}-\mu_{{B}}\right)^{2}$\\

$\widehat{x}_{i} \leftarrow \frac{x_{i}-\mu_{{B}}}{\sqrt{\sigma_{{B}}^{2}+\epsilon}}$\\

$y_{i} \leftarrow \gamma \widehat{x}_{i}+\beta \equiv \mathbf{B N}_{\gamma, \beta}\left(x_{i}\right)$\\

\caption{Batch Normalizing Transform \cite{ioffe2015batch}}
\label{alg:BN}
\Indp\Indpp
\end{algorithm}\DecMargin{1em}

%AAAAAAAAAAAAAAAAAAAAAAAAAAAAAAAAAAAAAAAAAAAA

For the read-out function ${R}$, we applied the mean operation on all node embeddings in the real graph $G$ and then applied a sigmoid activation function to compute the whole graph embeddings $\overline{s}$:  

\begin{equation}
\bar{s}=\sigma\left(\frac{1}{n} \sum_{i=1}^{n} h_{i}^{(L)}\right)
\label{eq:readout}
\end{equation}

In Algorithm \ref{alg:Inspection-L}, from line $7$ to $8$, as shown in Equation~\ref{eq:disc1} and Equation~\ref{eq:disc2}, for the discriminator ${D}$, we used a logistic sigmoid non-linear function to discriminate node embedding vector $\vec{h}_{i}$ against the real whole graph embedding $\overline{s}$ to calculate the score of $(\vec{h}_{i},\overline{s})$ being positive or negative:
\begin{equation}
{D}\left(h_{i}, \bar{s}\right)=\sigma\left(h_{i}^{T}{w} \bar{s}\right)
\label{eq:disc1}
\end{equation}

\begin{equation}
{D}\left(\widetilde{h}_{i}, \bar{s}\right)=\sigma\left(\widetilde{h}_{i}^{T}{w} \bar{s}\right)
\label{eq:disc2}
\end{equation}

 \begin{figure*}[!t]

    \centering
        \includegraphics[width=1.3\columnwidth]{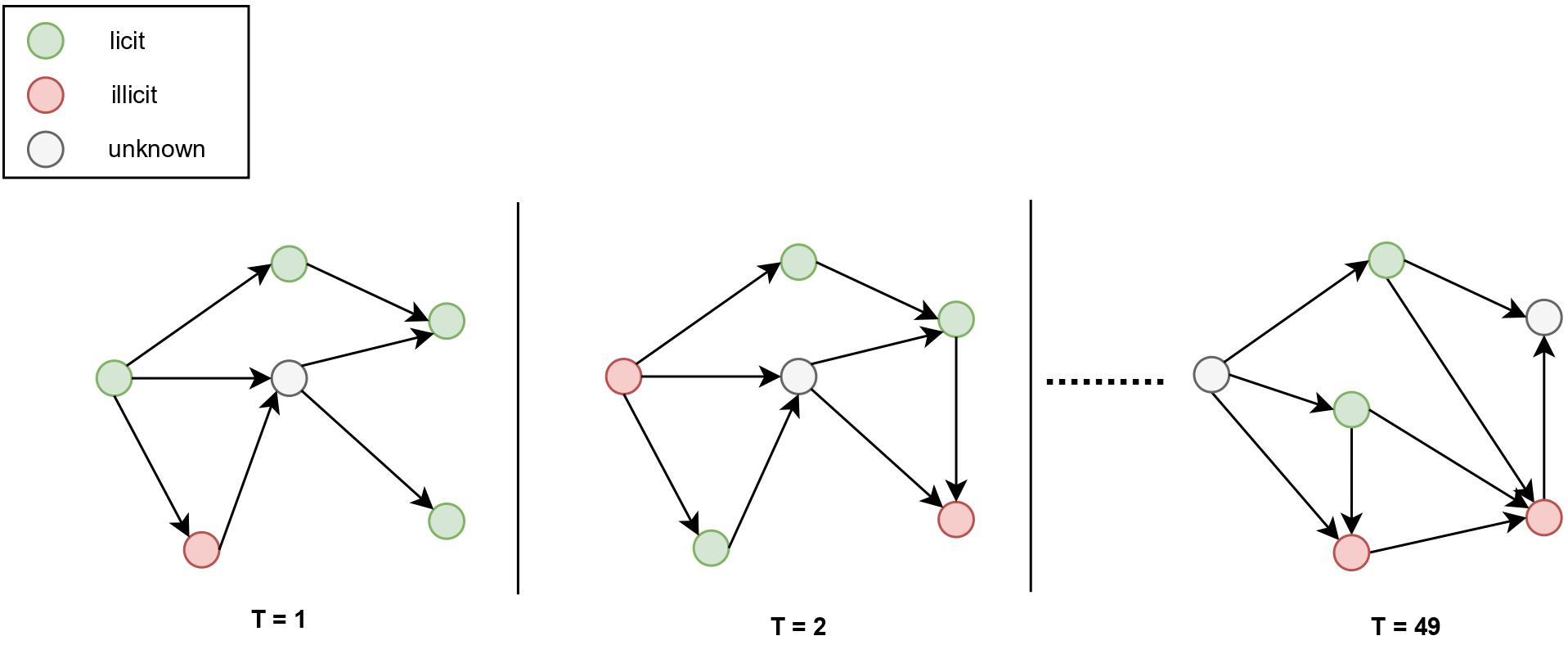}
        %\label{rfidtest_yaxis}
    \caption{Overview of Elliptic Dataset \cite{robinson} }
    \label{fig:dataset}
\end{figure*}

We then used a binary cross-entropy loss objective function (based on Equation~\ref{eq:dgi}, modified so that $N=M$) to perform gradient descent, as shown in Algorithm \ref{alg:Inspection-L}, line $10$. To perform gradient descent, we maximized the score if the node embedding is a true node embedding $\vec{h}_{i}$ and minimized the score if it is a corrupted node embedding $\vec{h}_{i}$ compared to the global graph summary generated by the read-out function ${R}$ (Equation~\ref{eq:readout}). As a result, we maximized the mutual information between patch representations and the whole real graph summary based on the binary cross-entropy loss function (BCE), as shown in Equation~\ref{eq:dgi} to perform gradient descent. After the training process, the trained encoder can be used to generate new graph embeddings for downstream purposes; in this case, the detection of illegal transactions.

 In our experiments, we used all 34 different Bitcoin transaction graphs to train the DGI with the GIN encoder in a self-supervised manner. For each training graph, we trained 300 epochs using an \textit{Adam} optimizer with a learning rate of 0.0001, as shown in Algorithm \ref{alg:Inspection-L}, line $10$.
 
\subsection{Supervised Machine Learning Classification}\label{Supervised Learning}
After the DGI training, we reused the encoder to generate node embeddings, as shown in Algorithm \ref{alg:Inspection-L}, line $11-12$ to train and test the RF classifier with 100 estimators. In our experiments, we performed 70:30 splitting, 34 different Bitcoin transaction graphs for training and the remaining 15 bitcoin transaction graphs for testing. All 34 training graphs were fed to DGI to train the GIN encoder in a self-supervised manner. Once the training phase was completed, we used a trained GIN encoder to extract all the node embeddings (all 34 graph node embeddings) in the training graphs. As the datasets consist of two labels, binary classification and unknown labels, we dropped unknown label data in the RF training and testing phases and only used label data for performance. We used all training graph node embeddings to train the RF in a supervised manner. For testing, we extracted the last 15 test graph node embeddings using the trained GIN and fed the node embeddings to the trained RF for illegal transaction detection.

%For training and testing the RF classifier, we used 100 estimators for training and testing. 
We experimented with the following three combinations of features and embeddings:
 
\begin{enumerate}
 \item \textbf{DNE : Node Embeddings only}: After the DGI training, we reused the encoder to generate node embeddings for training and testing the RF classifier, as mentioned above.
 
  \item \textbf{LF + DNE : Node Embeddings with LF features}: Similar to scenario 1, we also combined local features (i.e, first 94 raw features) with the node embeddings generated by the trained encode for training and testing the RF classifier, as mentioned above.
  
    \item \textbf{AF + DNE : Node Embeddings with AF Features}: Similar to scenario 1, we also combined all raw features (AF features) with the node embeddings generated by the trained encoder for training and testing the RF classifier, as mentioned above.
\end{enumerate}

 \subsection{Implementation Environments}\label{Supervised Learning}
 
  %TTTTTTTTTTTTTTTTTTTTTTTTTTTTTTTTTTTTTTTTTTTTTTTTTTTTTTT
\begin{table}[t]\small
\caption{Implementation environment specification}
\begin{tabular}{lll}
%  \hline
%  \multicolumn{4}{|c|}{Multiclass Classification Performance Comparison with state-of-the-arts Algorithms} \\
   \hline

  \textbf{Unit}  &   \textbf{Description} &  \\
  \hline
 Processor      & 2.3 GHz 2-core Inter Xeon(R) Processor\\

RAM  & 12GB \\ 

 GPU  &    Tesla P100 GPU 16GB \\ 
 
  Operating System  &    Linux \\ 
    Packages &    Sckit-learn, Numpy, Pandas, \\
             &    PyTorch Geometric, and Matplotlib \\
 \hline
\end{tabular}
\label{table:environment}
\end{table}
%TTTTTTTTTTTTTTTTTTTTTTTTTTTTTTTTTTTTTTTTTTTTTTTTTTT

Experiments were carried out using a $2.3 \mathrm{GHz}$ 2-core Intel(R) Xeon(R) processor with 12 GB memory and Tesla P100 GPU on a Linux operating system. The proposed approach was developed using the Python programming language with several statistical and visualization packages, such as Sckit-learn, Numpy, Pandas, PyTorch Geometric, and Matplotlib. Table \ref{table:environment} summarizes the system configuration.

\section{Experiments and Results}\label{Evaluation}

\subsection{Dataset}\label{Proposed}

% move the dataset after proposed method
In this paper, we adopted the Elliptic dataset \cite{AML}, which is the world's largest labeled dataset of bitcoin transactions. The Elliptic dataset \cite{AML} consists of 203,769 node as transactions and 234,355 directed transaction payment flows (i.e., transaction inputs, transaction outputs). The datasets also consists of 49 different timestep graphs, which are uniformly spaced with a two-week interval, as illustrated in \ref{fig:dataset}. Each connected transaction component consists of a time step that appears on the blockchain in less than three hours. Our $G$ represents one such transaction graph for the 49.

In the Elliptic dataset \cite{AML}, 21\% of the node entities are labeled as licit, and only 2\% are labeled as illicit. The remaining node entities are unlabeled but have node features. These node entities consist of 166 features (AF features), among which the first 94 features contain local information (LF features) of the transactions, including the time step, transaction fees, and the number of inputs or outputs. The remaining 72 features are aggregated features. These features can be obtained by aggregating transaction information from one-hop backward/forward graph nodes, such as the standard deviation, minimum, maximum, and correlation coefficients of the neighbor transactions for the same information data. More importantly, all features were obtained using only publicly available information.

\subsection{Performance Metric}\label{Experimental Results}

To evaluate the performance of the proposed methods, the standard metrics listed in Table~\ref{tab: metrics} were used, where $TP$, $TN$, $FP$ and $FN$ represent the number of True Positives, True Negatives, False Positives and False Negatives, respectively.

\begin{figure*}[t]
 \centering
        \includegraphics[height=7.5cm,width=1.95\columnwidth]{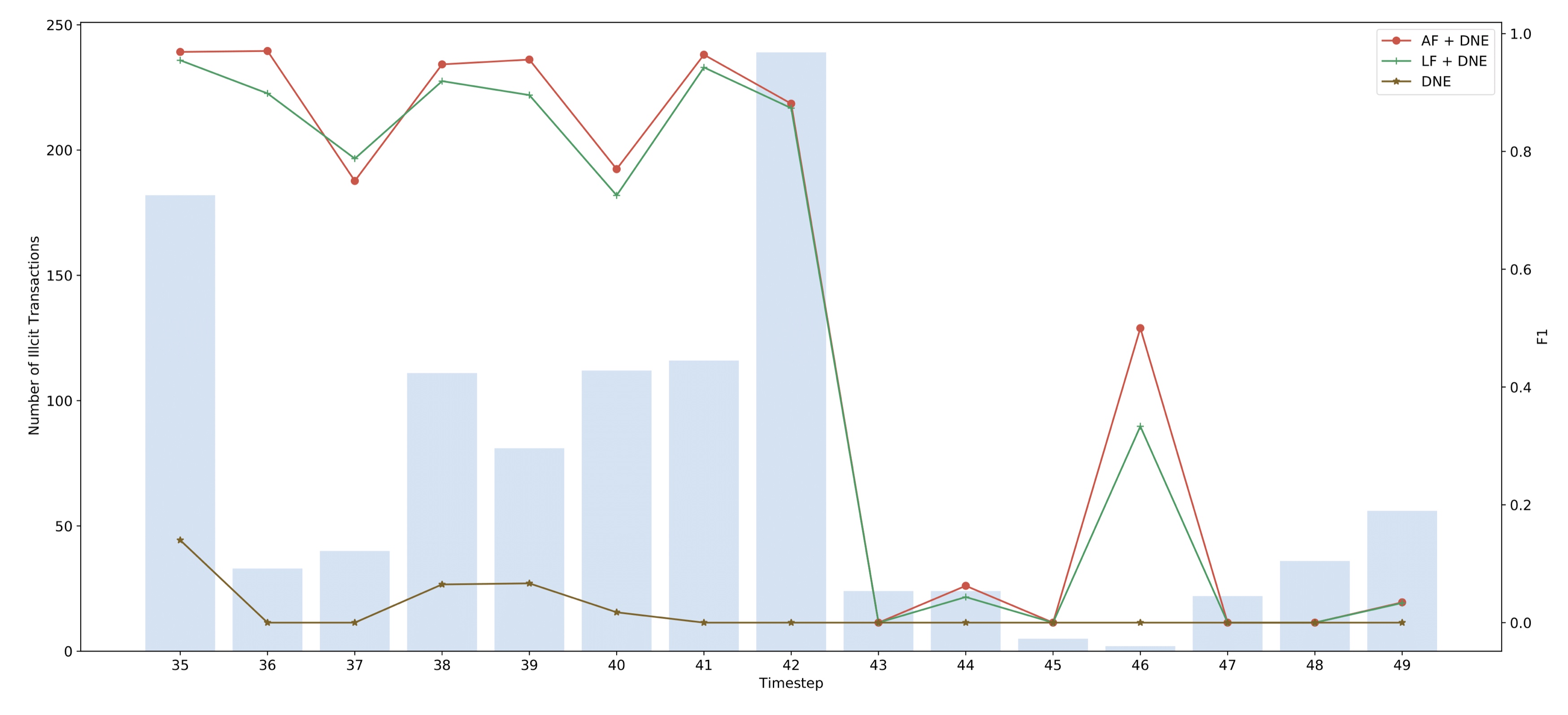}
        %\label{rfidtest_yaxis}
    \caption{Illicit F1 over test timestep}
    \label{fig:recall}
\end{figure*}

%TTTTTTTTTTTTTTTTTTTTTTTTTTTTTTTTTTTTTTTTTTT
\begin{table}[!t]\small
\renewcommand{\arraystretch}{1.4}
\caption{Evaluation metrics used in this study}
\centering
\begin{tabular}
{|c|c|} \hline
\textbf{Metric} & \textbf{Definition} \\ \hline 
\small{ Detection Rate (Recall)} &  $\frac{TP}{TP+FN}$ \\ \hline
\small{Precision} & $\frac{TP}{TP+FP}$ \\ \hline
\small{F1-Score} & $ 2 \times \frac{Recall\times Precision}{Recall + Precision}$ \\ \hline

\small{AUC-Score} & $\int_{0}^{1} \frac{T P}{T P+F N} d \frac{F P}{T N+F P}$ \\ \hline

\end{tabular}
\label{tab: metrics}
\end{table}
%TTTTTTTTTTTTTTTTTTTTTTTTTTTTTTTTTTTTTTT

In Table \ref{tab: metrics}, true positive (TP) denotes the total number of true positives, true negative (TN) indicates the total number of false positives, false positive (FP) denotes the total number of false negatives and false negative (TN) shows the total number of true negatives. 
The proposed method was evaluated using Precision, Recall, F1-score and Area under the receiver operating characteristics (ROC) curve. All the above metrics can be obtained using the confusion matrix (CM).

Accuracy indicates that the model is well learned in case of a balanced test dataset; however, for imbalanced scenarios, as in this case, only considering accuracy measures may lead to misleading conclusion,s since it is strongly biased in favor of the licit majority class. Thus, for this case, recall and F1-score metrics provide a more reasonable explanation of the model's performance.

Recall (also known as Detection Rate) is the total number of true positives divided by the total number of true positives and false negatives. If the recall rate is very low, this means that the classifier cannot detect illicit transactions.

Precision measures the quality of the correct predictions. This is the number of true positives divided by the number of true positives and false positives. If the false positive is very high, it will cause low precision. Our goal is to maximize the precision as much as possible.

F1-score is the trade-off between precision and recall. Mathematically, it is the harmonic mean of precision and recall.

The area under the curve (AUC) computes the trade-off between sensitivity and specificity, plotted based on the trade-off between the true positive rate on the y-axis and the false positive rate on the x-axis. Our goal is to maximize the AUC score as much as possible, making is closer to 1.0.

 \subsection{Experimental Results}\label{Experimental Results}

 \begin{table}[t]
\centering
\caption{Results of binary classification by Inspection-L compared to the state-of-the-art. AF refers to all raw features, LF refers to the local raw features, i.e., the first 94 raw features, GNE refers to the node embeddings generated by GCN in \cite{AML} using labels and DNE refers to the node embeddings computed by DGI without using labels. }
\scalebox{0.8}{
\begin{tabular}{|p{14em}|ccc|c|}
% \begin{tabular}{lrrrr}
% \cline{2-4}
\hline
& \multicolumn{3}{c|}{ \textbf{Illicit}} & \\ 
\multicolumn{1}{|c|}{\textbf{Method}} &
\multicolumn{1}{c}{\textbf{Precision}} &
\multicolumn{1}{c}{\textbf{Recall}} & 
\multicolumn{1}{c|}{\textbf{F1}} & \multicolumn{1}{c|}{\textbf{AUC}}\\ \hline
\multicolumn{1}{|l|}{ Logistic Regr\textsuperscript{AF} \cite{AML} } &  0.404 & 0.593 & 0.481 & $-$ \\
\multicolumn{1}{|l|}{ Logistic Regr \textsuperscript{AF + GNE} \cite{AML} }   & 0.537 & 0.528 & 0.533  & $-$ \\
\multicolumn{1}{|l|}{ Logistic Regr \textsuperscript{LF} \cite{AML} }   & 0.348 & 0.668 &  0.457 & $-$\\
\multicolumn{1}{|l|}{ Logistic Regr \textsuperscript{LF + GNE} \cite{AML} }   & 0.518 & 0.571 & 0.543 & $-$ \\
\multicolumn{1}{|l|}{ RandomForest \textsuperscript{AF} \cite{AML} }   & 0.956 & 0.670 &  0.788 & $-$\\
\multicolumn{1}{|l|}{ RandomForest \textsuperscript{AF + GNE} \cite{AML} }   & 0.971 & 0.675 & 0.796 & $-$\\

\multicolumn{1}{|l|}{ RandomForest \textsuperscript{AF } \cite{vassallo2021application} }   & 0.897	& 0.721 & 0.800 & $-$\\

\multicolumn{1}{|l|}{ RandomForest \textsuperscript{AF + GNE} \cite{vassallo2021application} }   & 0.958& 0.715 & \textbf{0.819} & $-$\\

\multicolumn{1}{|l|}{ XGB \textsuperscript{AF } \cite{vassallo2021application} }   & 0.921	& 0.732 & 0.815 & $-$\\

\multicolumn{1}{|l|}{ XGB \textsuperscript{AF + GNE} \cite{vassallo2021application} }   & 0.986 & 0.692 & 0.813 & $-$\\

\multicolumn{1}{|l|}{ RandomForest \textsuperscript{LF} \cite{AML} }   & 0.803 & 0.611 &  0.694 & $-$\\
\multicolumn{1}{|l|}{ RandomForest \textsuperscript{LF + GNE} \cite{AML} }   & 0.878 & 0.668 & 0.759 & $-$\\
\multicolumn{1}{|l|}{ MLP \textsuperscript{AF} \cite{AML} }   & 0.694 & 0.617 & 0.653 & $-$\\
\multicolumn{1}{|l|}{ MLP \textsuperscript{AF + GNE} \cite{AML} }   & 0.780 & 0.617 & 0.689 & $-$ \\
\multicolumn{1}{|l|}{ MLP \textsuperscript{LF} \cite{AML} }   & 0.637 & 0.662 & 0.649 & $-$\\
\multicolumn{1}{|l|}{ MLP \textsuperscript{LF + GNE} \cite{AML}}   & 0.681 & 0.578 &  0.625 & $-$ \\
\multicolumn{1}{|l|}{ GCN \cite{AML}}   & 0.812 & 0.512 & 0.628  & $-$\\
\multicolumn{1}{|l|}{ GCN \cite{Alarab}}   & 0.899 & 0.678 & 0.773 & $-$\\

\multicolumn{1}{|l|}{ Skip-GCN \cite{AML}}   & 0.812 & 0.623 &  0.705 & $-$ \\
\multicolumn{1}{|l|}{ EvolveGCN \cite{AML}}   & 0.850 & 0.624 & 0.720 & $-$\\
\hline

% heighted 

\multicolumn{1}{|l|}{Inspection-L \textsuperscript{DNE} (RF)}   & 0.593  & 0.032 & 0.061  & 0.735\\
\multicolumn{1}{|l|}{Inspection-L \textsuperscript{LF + DNE} (RF)}   & 0.906  & 0.712 & 0.797  & 0.895 \\
\multicolumn{1}{|l|}{\textbf{Inspection-L \textsuperscript{AF + DNE} (RF)}}   & \textbf{0.972}  & \textbf{0.721} & \textbf{0.828}  & \textbf{0.916}  \\
\hline

\end{tabular}
}

\label{tab:elicipt}

\end{table}

\begin{figure*}[t!]
  \centering
   \begin{subfigure}[b]{0.25\textwidth}
    \centering
    \includegraphics[width = \linewidth]{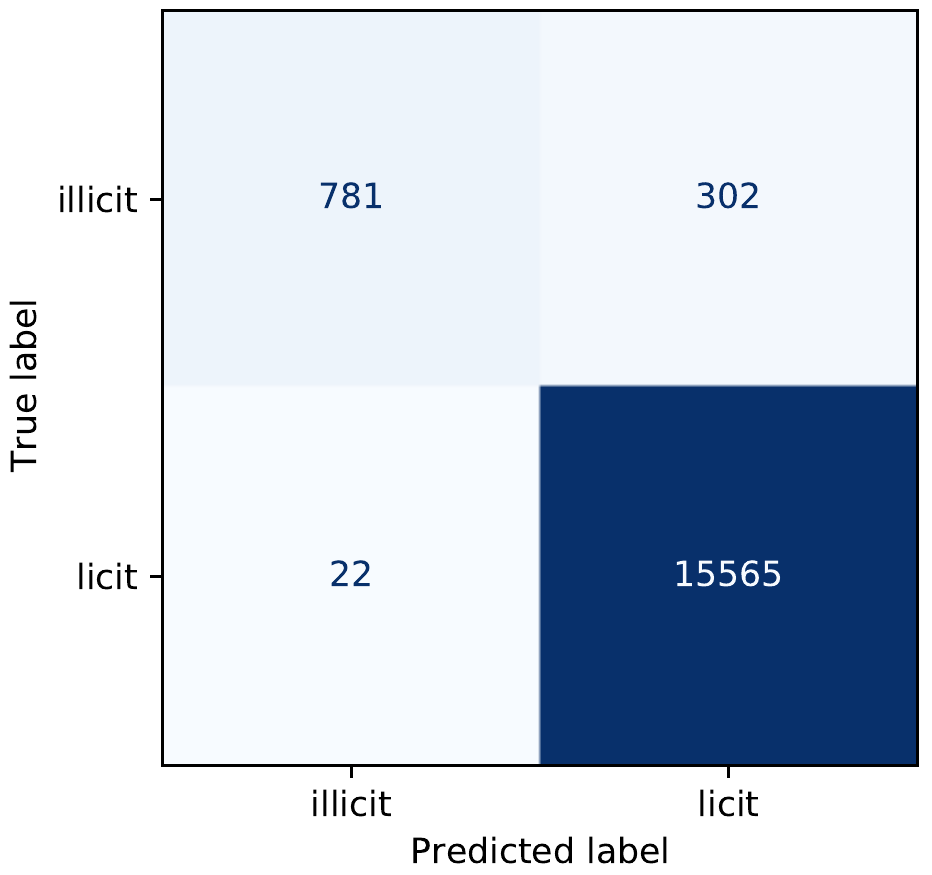}
    \caption{AF $+$ DNE}
    \label{fig:DNE_AF}

  \end{subfigure}%
  \hspace{1em}% Space between image A and B
   \begin{subfigure}[b]{0.25\textwidth}
    \centering
    \includegraphics[width = \linewidth]{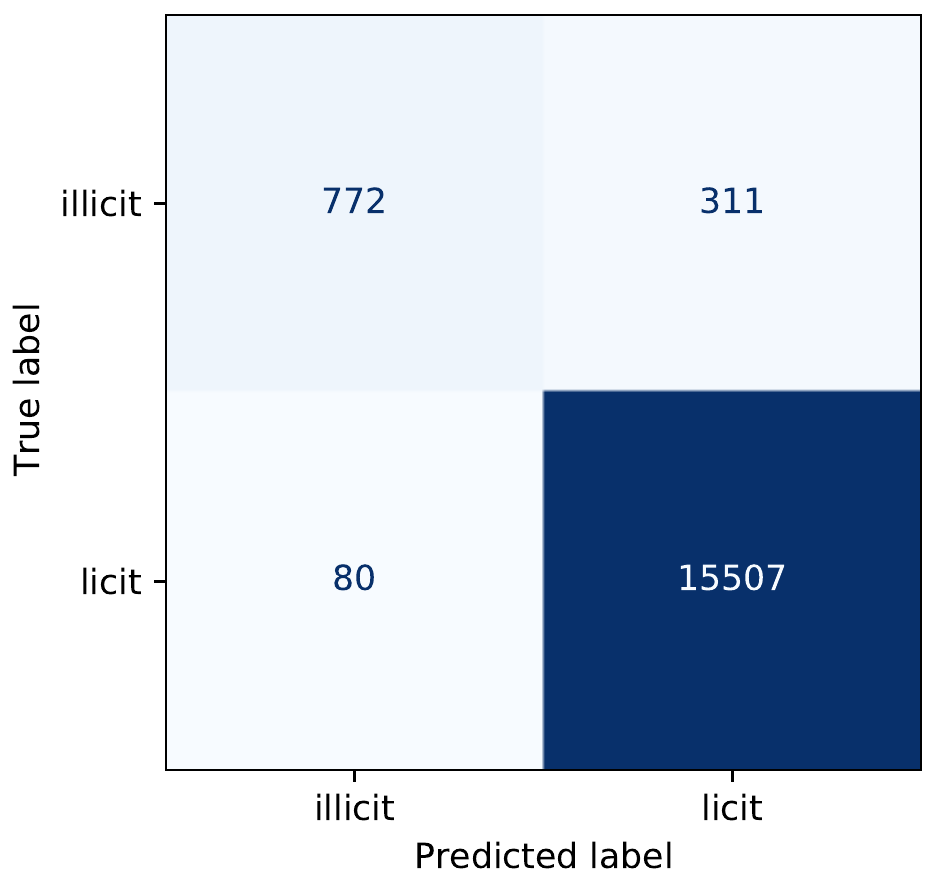}
    \caption{LF $+$ DNE}
      \label{fig:DNE_LF}

  \end{subfigure}%
  \hspace{2em}% Space between image B and C
  \begin{subfigure}[b]{0.252\textwidth}
    \centering
    \includegraphics[width = \linewidth]{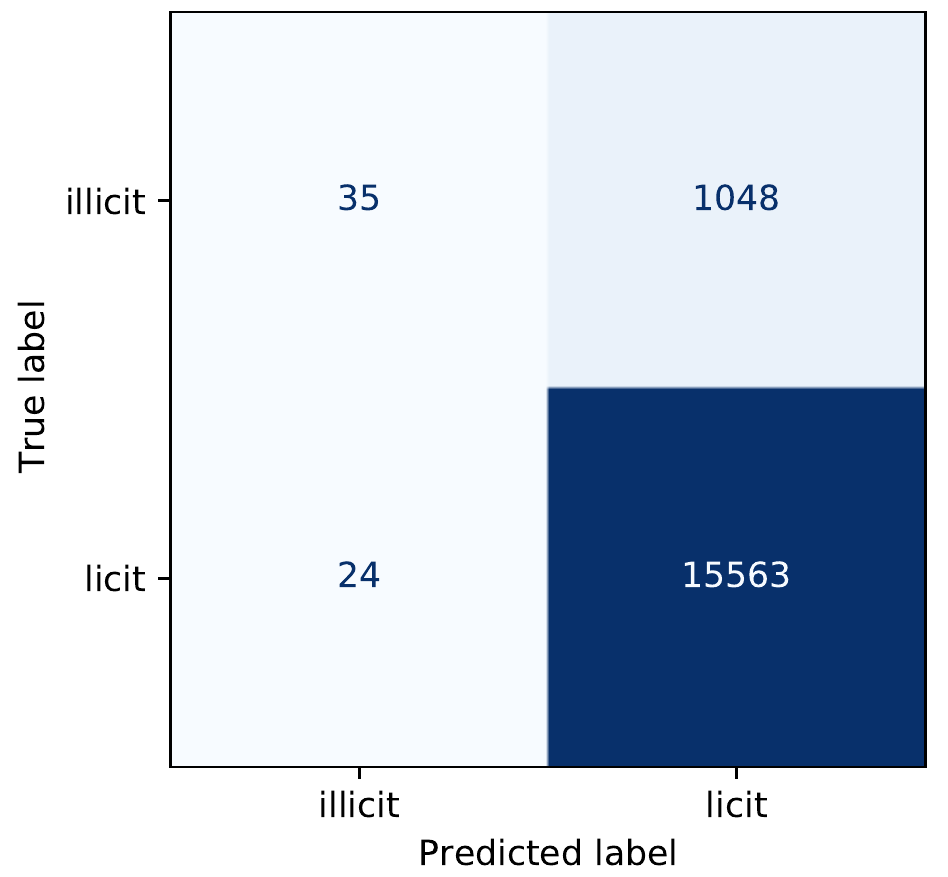}
    \caption{DNE}
  \label{fig:DNE_conf}
  \end{subfigure}
  \caption{Confusion Matrix}
  \label{fig:confusion}
\end{figure*}

Table \ref{tab:elicipt} shows the corresponding results of our \mbox{Inspection-L} compared to the state-of-the-art in terms of the key metrics. 
As can be observed from the table, regarding to illicit F1-Score, Inspection-L (LF+DNE and AF+DNE) outperforms the best reported classifiers. 
In the best-performing variant, AF+DNE, we concatenated the node embeddings generated from DGI with all original raw features (AF).
The experiment achieved an F1 score and Recall of 0.828 and 0.721, respectively. Using all features (AF) with node embeddings (DNE) as input for classification, the ML model's performance significantly increased, with an AUC of 0.916, compared to 0.735 when only the node embeddings were used for classification. The experiments demonstrate that graph information (node embeddings) is useful to enhance the transaction representations (embeddings).

In the second experiment LF+DNE, we concatenated the node embeddings generated from DGI with the local features (LF), which can achieve an F1-score and Recall of 0.712 and 0.797, respectively. Both  the results were superior to the state-of-the-art algorithms. 

These results demonstrate the ability of our self-supervised GIN-based approach to generate an enhanced feature set to improve anti-money-laundering detection performance. Furthermore, the results show that the accuracy of the model improves with the enhanced feature set, which contains summary information. Note that the summary information in the AF feature set consists of 1-hop forward and 1-hop backward neighborhood summaries for each node. Unfortunately, the Elliptic dataset does not provide detailed information regarding the feature descriptions, possibly due to confidentially reasons, which limits our ability to provide a deeper discussion.
%Our RF classifier, trained using all raw features and embedding features, not only detect illicit transactions, but can also has a a low false alarm rate.

Figure \ref{fig:recall} shows the F1 measure of the three different model variants across various testing timesteps. Interestingly, none of the three variants can detect new illicit transactions with high precision after dark market shutdown, which occurs at time step 43 \cite{AML}. Thus, we note that developing robust methods to detect illicit transactions without their being affected by emerging events is a major challenge that future works need to address.

Figure \ref{fig:confusion} shows the confusion matrix of the three different scenarios. Although the classifier trained with embedding features cannot accurately detect illicit transactions, it rarely classifies licit transactions as illicit. Therefore, the false alarm rate is very low, as shown in Figure \ref{fig:DNE_conf}. The RF classifier trained using both raw features and embedding features, shown in Figure \ref{fig:DNE_AF},\ref{fig:DNE_LF}, has the advantage of achieving a high detection rate and a low false alarm rate. As a result, the experimental results demonstrate that DNE node embeddings can be used for feature augmentation to improve overall detection performance. 

\subsection{Broader applications of AML}
The blockchain operates as a decentralized bank for bitcoin cryptocurrency \cite{nakamoto2008bitcoin}.
All bitcoin transactions are permanently recorded on the blockchain, which is a visible and verifiable public ledger \cite{van2018bitcoin}.
Bitcoin addresses are not registered to individuals, in contrast to bank accounts \cite{kshetri2017crypto}.
Thus, due to this pseudo-anonymity \cite{hu2019characterizing}, bitcoin and other crypto-currencies are increasingly used for ransomware \cite{kshetri2017crypto}, ponzi schemes \cite{hu2019characterizing} and illicit material trade on the dark web \cite{kulatilleke2021fdgatii}

While bitcoin transactions are difficult to track, they are not completely anonymous \cite{kshetri2017crypto}. 
Users can be traced by their IP addresses and transaction flows \cite{van2018bitcoin}. An analysis of the bitcoin graph can reveal suspicious behavior patterns characteristic of money laundering \cite{kshetri2017crypto}.
To break the tell-tale transnational link between bitcoin transactions and illegal activity, bitcoin mixing services provide a new, untainted bitcoin address from their reserves and the pay-outs are spread out over time \cite{kshetri2017crypto}. Bitcoin Fog is a service that hides transaction origins by bundling multiple inputs into a smaller number of larger outputs \cite{hu2019characterizing}. 
However, the additional obscuring activities themselves could add characteristic signatures into transaction flows.
Thus, it is still possible to detect patterns in the underlying transaction flow to facilitate AML detection \cite{hu2019characterizing,AML}.
Unfortunately, next-generation cryptocurrencies such as Monero, Dash, and Z-Cash, with built-in anonymity features, make tracking and detection challenging \cite{kshetri2017crypto}. As a result, there is a constant need for improved AML detection methodologies.

\section{Conclusions and Future Work}\label{Conclusions}
This paper presents a novel approach for the detection of illicit Bitcoin transactions based on self-supervised GNNs. We first used the DGI to generate the node embedding with raw features to train the Random Forest for detection. Our experimental evaluation indicates that our approach performs exceptionally well and outperforms the state-of-the-art ML-based/Graph-based classifier overall. The evaluation results of our initial classifier demonstrate the potential of using a self-supervised GNN-based approach for illegal transaction detection in cryptocurrencies. We hope to inspire others to work on the important challenge of using graph machine learning to perform financial forensics through this research, which is lacking in the current research. In the future, we plan to integrate this with unsupervised anomaly detection algorithms to detect illegal transactions in an unsupervised manner.

\bibliography{main}

\end{document}